\documentclass[conference]{IEEEtran}
\IEEEoverridecommandlockouts
\usepackage{cite}
\usepackage{amsmath,amssymb,amsfonts}
\usepackage{algorithmic}
\usepackage{graphicx}
\usepackage{textcomp}
\usepackage{xcolor}
\usepackage{pifont}
\usepackage{float}
\usepackage{url}
\usepackage{graphicx}
\usepackage{subcaption}
\usepackage{enumerate}
\usepackage{tabularx} % add this in your preamble

\usepackage{subcaption}

\usepackage[utf8]{inputenc}
\usepackage{newunicodechar}
\newunicodechar{≈}{\approx}

\usepackage{hyperref}
\hypersetup{
    colorlinks=true,
    allcolors=blue
}

\usepackage{afterpage}
\usepackage{booktabs}
\usepackage{tabularray}
\usepackage{microtype}
\usepackage{multirow}
\usepackage{array}
\newcommand{\PreserveBackslash}[1]{\let\temp=\\#1\let\\=\temp}
\newcolumntype{C}[1]{>{\PreserveBackslash\centering}p{#1}}
\newcolumntype{R}[1]{>{\PreserveBackslash\raggedleft}p{#1}}
\newcolumntype{L}[1]{>{\PreserveBackslash\raggedright}p{#1}}
\usepackage{booktabs}
\newcommand{\modelName}{SpectroFusion-ViT}
\newcommand{\backbone}{EfficientViT-b0}
\newcommand{\etal}{\textit{et al.}}
\def\BibTeX{{\rm B\kern-.05em{\sc i\kern-.025em b}\kern-.08em
    T\kern-.1667em\lower.7ex\hbox{E}\kern-.125emX}}
\begin{document}

\title{SpectroFusion-ViT: A Lightweight Transformer for Speech Emotion Recognition Using Harmonic Mel–Chroma Fusion}
% Speech Emotion Recognition via Acoustic Feature Transformation through Harmonic Mel Spectrum Fusion}

%\author{(Anonymous \confName  submission)}

\author{ 
    \IEEEauthorblockN{
    Faria Ahmed,
    Rafi Hassan Chowdhury, 
    Fatema Tuz Zohora Moon,
    Sabbir Ahmed\\}

    \IEEEauthorblockA{
        Department of Computer Science and Engineering\\
        Islamic University of Technology, Gazipur 1704, Bangladesh}

    \IEEEauthorblockA{\{fariaahmed, rafihassan, fatema, sabbirahmed\}@iut-dhaka.edu
    } 
}
%\author{Anonymous ICCIT Submission}

%%%%%%% ADDING CONFERENCE info in TOP-LEFT Corner %%%%%%%%%%
%  Source: https://tex.stackexchange.com/questions/561793/how-to-get-ieee-conference-template-to-show-conference-name-in-header

\makeatletter
\let\old@ps@IEEEtitlepagestyle\ps@IEEEtitlepagestyle
\def\confheader#1{%
    % for the first page
    \def\ps@IEEEtitlepagestyle{%
        \old@ps@IEEEtitlepagestyle%
        \def\@oddhead{\strut\hfill#1\hfill\strut}%
        \def\@evenhead{\strut\hfill#1\hfill\strut}%
    }%
    \ps@headings%
}

\makeatother

\confheader{
        \parbox{20cm}{2025 28th International Conference on Computer and Information Technology (ICCIT)\\
        19-21 December 2025, Cox’s Bazar, Bangladesh}
}

\IEEEpubid{
\begin{minipage}[t]{\textwidth}\ \\[10pt]
       \small{979-8-3315-7867-1/25/\$31.00 \copyright2025 IEEE }
\end{minipage}
}
%%%%%%%%% FOR ARXIV %%%%%%%%%%%
%\IEEEpubid{
%\begin{minipage}[t]{\textwidth}\ \\[10pt]
%      \small{Accepted in 28th ICCIT \copyright2025 IEEE }  
%\end{minipage}
% }
\maketitle

\begin{abstract}

Speech is a natural means of conveying emotions, making it an effective method for understanding and representing human feelings. Reliable speech emotion recognition (SER) is central to applications in human–computer interaction, healthcare, education, and customer service. However, most SER methods depend on heavy backbone models or hand-crafted features that fail to balance accuracy and efficiency, particularly for low-resource languages like Bangla.
In this work, we present \modelName, a lightweight SER framework built utilizing \backbone, a compact Vision Transformer architecture equipped with self-attention to capture long-range temporal and spectral patterns. The model contains only 2.04M parameters and requires 0.1 GFLOPs, enabling deployment in resource-constrained settings without compromising accuracy. 
Our pipeline first performs preprocessing and augmentation on raw audio, then extracts Chroma and Mel-frequency cepstral coefficient (MFCC) features. These representations are fused into a complementary time–frequency descriptor that preserves both fine-grained spectral detail and broader harmonic structure. Using transfer learning, \backbone~is fine-tuned for multi-class emotion classification.
We evaluate the system on two benchmark Bangla emotional speech datasets, SUBESCO and BanglaSER, which vary in speaker diversity, recording conditions, and acoustic characteristics. The proposed approach achieves 92.56\% accuracy on SUBESCO and 82.19\% on BanglaSER, surpassing existing state-of-the-art methods. 
These findings demonstrate that lightweight transformer architectures can deliver robust SER performance while remaining computationally efficient for real-world deployment.

% (The codes and models will be made publicly available upon acceptance.)

\end{abstract}


\section{Introduction}

Speech is the primary mode of communication for humans, and it naturally conveys emotional information through variations in prosody, pitch, and tone. Advanced context-aware artificial intelligence systems rely on accurate speech emotion recognition (SER) to better interpret human intentions and emotional states \cite{cummins2019review}. %karam2020suicide
% As a result, SER plays a critical role in human–computer interaction \cite{zhou2022robotaffect}, healthcare \cite{chahar2023emotionaware}, education \cite{kaur2022multimodal}, customer service \cite{zhang2019sercallcenter}, etc \cite{dar2024databases}.
As a result, SER plays a critical role in human–computer interaction, healthcare, education, customer service, and so on \cite{wani2021speech}.
% and emotion-aware caregiving \cite{inkster2018digitalmentalhealth}. 
Systems capable of reliably identifying emotions from natural speech can enable more intuitive, adaptive, and personalized interactions \cite{dmello2014affectivesensitive, aziz2023banglaSER}. %

Early work in affective computing commonly used statistical classifiers such as Support Vector Machines (SVMs) and Hidden Markov Models (HMMs) \cite{chakraborty2022phase, 11013930}. However, these methods are highly dependent on hand-crafted acoustic features, which often fail to capture the complex temporal and spectral patterns present in emotional speech. Modern deep learning (DL) architectures overcome these limitations through automatic feature learning, enabling state-of-the-art performance across a wide range of downstream tasks \cite{rafi2025mangoLeafVit, ivan2024meme,11022513,11022615, khan-etal-2023-banglachq}.
A typical pipeline for speech understanding extracts acoustic features or transforms raw audio into spectrogram-like images using the Short-Time Fourier Transform (STFT) \cite{yadav2021stftmfcc}, followed by convolutional neural networks (CNNs), such as ResNet variants \cite{he2016deepresiduallearning}. Recent studies introduced attention mechanisms, including LSTM models with self-attention and attention-enhanced CNNs, which reported competitive performance \cite{9654185, chakraborty2022phase, yasmeen2021csvcNet, khan2022rethinking, hasan2023GaitGCN}. More advanced multimodal frameworks- integrating speech, facial cues, or text—have achieved even higher accuracy \cite{nithyasri2024multimodal}.

Despite these advances, several challenges still limit the practicality and scalability of existing SER systems. Many modern SER models rely on heavy backbone architectures with millions of parameters, making real-time deployment difficult. Hand-crafted features often fail to capture subtle spectral or temporal nuances in emotional speech, while multimodal systems require extra sensors or hardware. In addition, most SER research focuses on widely spoken languages, limiting generalization to less-resourced linguistic settings such as Bangla. These limitations highlight the need for a lightweight, efficient, and linguistically adaptable SER framework that balances accuracy with computational efficiency.

In this connection, we propose \modelName~a transformer-based framework that leverages \backbone~to model long-range temporal and spectral dependencies in speech. Raw audio is amplitude-normalized and on-the-fly augmented before acoustic features are extracted. Chroma and Mel-frequency cepstral coefficients (MFCCs) are combined into a fused time–frequency descriptor that retains complementary spectral properties. %This representation is used to fine-tune an \backbone backbone \cite{cai2024efficientvitmultiscalelinearattention}. 
We evaluate using 5-fold cross-validation on the SUBESCO and BanglaSER corpora and compare their performance against strong CNN and ViT baselines.
In summary, this work makes the following key contributions:
\begin{enumerate}[i.]
    \item A lightweight Bangla SER framework based on \backbone, offering only 2.04M parameters and 0.1 GFLOPs while maintaining competitive accuracy.
    \item  A tailored fusion of Chroma and MFCC features designed for transformer-based architectures, addressing the limitations of prior Bangla SER approaches that rely primarily on MFCC or other hand-crafted features.
    \item A comprehensive augmentation pipeline that enhances model generalization in low-resource Bangla emotional speech settings.
\end{enumerate}

% These contributions enable our system to achieve state-of-the-art performance on the SUBESCO and BanglaSER datasets, while remaining computationally efficient for real-world deployment.

%%%%%%%%%%%%%%%%%%%%%%%%%%%%%%%%%
\section{Literature Review}

Speech Emotion Recognition (SER) has gained significant research interest due to its wide range of applications. Despite progress, achieving robust and high-accuracy emotion recognition remains challenging due to the variability in speakers, languages, and recording conditions \cite{wani2021speech}. These factors make emotional cues difficult to model reliably, especially in low-resource linguistic contexts.

Prior works in SER have often reported suboptimal performance, particularly when relying on conventional architectures or limited feature representations \cite{latif2023survey}. Siddiqui \etal \cite{siddiqui2019autoencoder} evaluated deep autoencoder networks on emotional speech from both neurotypical and autistic children, obtaining accuracies as low as 33.3\% for autistic speech. This highlights the difficulty of generalizing across diverse speaker groups. Another work introduced a multi-feature, multi-lingual fusion approach, which improved performance but still remained below 70\% on well-established SER datasets \cite{wang2020multi}. Bangla SER research reflects similar challenges. Ayon \etal \cite{ayon2022bangla} reported accuracies as low as 51.33\% using recurrent neural networks for six-emotion classification, with an ensemble method reaching about 70\%. Islam \etal \cite{islam2023bangla} achieved 77–80\% accuracy using deep learning models. These results demonstrate the need for more expressive model architectures and richer feature representations tailored specifically for Bangla emotional speech.

Frequency-based representations such as STFT and MFCC have been widely studied across speech classification tasks. Yadav and Ranjan~\cite{yadav2021stftmfcc} compared the effectiveness of STFT and MFCC features for lie detection and found that MFCCs outperform STFT, emphasizing how feature effectiveness depends heavily on the task. MFCCs have also shown strong performance in speaker recognition~\cite{speakerrec2023mfcc}, yet transferring these features directly to SER is nontrivial because emotional cues encode more subtle spectral patterns. Composite feature strategies have been explored in music analysis, where combined MFCC–STFT representations have improved genre and instrument classification~\cite{musicgenre2021}, demonstrating that complementary spectral cues can enhance discriminability. However, such fusion strategies remain underexplored in SER, particularly for languages with limited annotated data.

Collectively, these observations highlight that improving SER accuracy beyond established benchmarks, particularly on specialized datasets like BanglaSER remains difficult. The inconsistent results across related tasks suggest that traditional spectral features alone may be insufficient for modeling nuanced emotional cues. At the same time, frequency-information-driven methods have demonstrated promising gains in other audio classification domains, motivating the exploration of more expressive spectral fusion strategies. These gaps underscore the opportunity for lightweight transformer-based architectures, such as \modelName, combined with richer harmonic feature integration to advance SER performance.

\section{Methodology} \label{Methodology}

\begin{figure*}[!t]
    \centering
    \includegraphics[width=.9\linewidth]{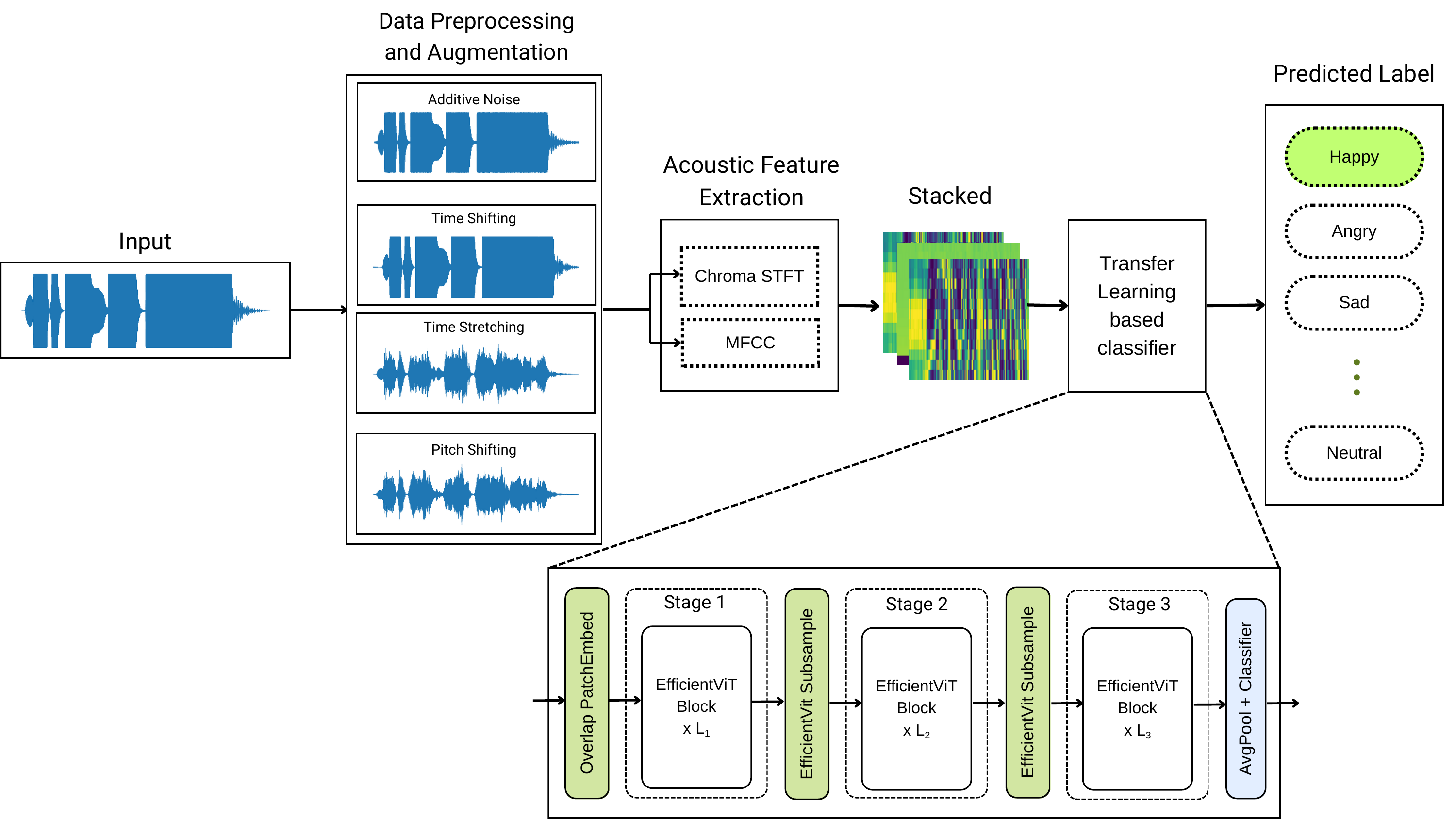}
    \caption{Overview of the proposed \modelName~framework for speech emotion recognition. Raw audio is first normalized and augmented, after which Chroma and MFCC features are extracted. These complementary representations are fused into a joint time–frequency descriptor and fed into the \backbone~backbone for transfer learning–based emotion classification.}
    \label{fig:overview_of_proposed_framework}
\end{figure*}

The proposed \modelName~framework, illustrated in \figureautorefname~\ref{fig:overview_of_proposed_framework}, takes real-time audio input and predicts the corresponding emotion label. The pipeline begins with preprocessing, where the raw waveform is normalized, augmented, and transformed into acoustic features. These features are then passed to a lightweight Vision Transformer backbone fine-tuned for emotion classification. To ensure robust evaluation and mitigate overfitting, we adopt a 5-fold cross-validation protocol in which three folds are used for training, one for validation, and one for testing. Averaging results across the five folds provides a more reliable estimate of model performance while reducing bias associated with any single train–test split.

\subsection{Dataset}

\subsubsection{SUBESCO Dataset}

The SUST Bangla Emotional Speech Corpus (SUBESCO) \cite{shahin2021subesco} is one of the largest and most rigorously validated emotional speech datasets available for Bangla. It contains approximately 7,000 utterances ($\approx$7 h 40 m) recorded by 20 professional speakers balanced by gender (10 male, 10 female). Each speaker produced 10 fixed Bangla sentences across seven emotional categories—anger, disgust, fear, happiness, neutral, sadness, and surprise—with five repetitions per sentence. The average duration of each recording is about 4 seconds (Details in \tableautorefname~\ref{table:subesco_summary}).

To ensure perceptual reliability, the dataset was validated by 50 human evaluators. Each utterance, except those labeled as `disgust', was rated four times by both male and female raters. The mean recognition accuracy exceeded 70\% across emotions, indicating strong annotation quality and making SUBESCO a robust benchmark for Bangla SER research.

\begin{table}[!h]
\centering
\caption{Properties of the SUBESCO dataset \cite{shahin2021subesco}}
\label{table:subesco_summary}
\begin{tabular}{l r}
\toprule
\textbf{Property} & \textbf{Value} \\
\midrule
Speakers & 20 (10 male, 10 female) \\
Sentences per speaker & 10 \\
Emotional classes & 7 (anger, disgust, fear, \\
& happiness, neutral, sadness, surprise) \\
Repetitions per sentence & 5 \\
Total utterances & 7,000 \\
Avg. utterance duration & $\approx$ 4 s \\
Total corpus duration & $\approx$ 7 h 40 m \\
Human evaluators & 50 \\
Avg. recognition accuracy & $>$70\% \\
\bottomrule
\end{tabular}
\end{table}

\subsubsection{BanglaSER}
The Bangla Speech Emotion Recognition (BanglaSER) dataset is a publicly available emotional speech corpus designed specifically for Bangla SER research. It contains recordings from 34 speakers (17 male and 17 female) between 19 and 47 years of age. The audio samples were captured in natural, real-world environments using smartphones and laptops, introducing variability in background noise and recording conditions. Each participant uttered three fixed Bangla sentences, repeated three times across five emotional categories—angry, happy, sad, surprise, and neutral—resulting in a total of 1,467 utterances. Each class contains 306 samples, except the neutral category, which includes 243 samples.

All recordings are approximately 3–4 seconds long and stored in 16-bit PCM WAV format. To ensure the reliability of annotation, human evaluators performed perceptual validation, achieving an average recognition accuracy of 80.5\%. A summary of the dataset properties is provided in Table~\ref{table:banglaser_properties}.

\subsection{Preprocessing and Augmentation}

Each recording was amplitude-normalized and augmented online during training to reduce overfitting. Each waveform was expanded into six variants, consisting of the original signal and multiple perturbed versions, while validation and test sets were kept unchanged. The six variants include samples produced via additive noise with a noise factor of 0.005, time stretching with rates of 1.1 \& 1.2, time shifting with a maximum value of 0.2, pitch shifting by 2 \& 4 step, and the original sample. Augmentation was applied after dataset partitioning in a controlled manner to prevent data leakage \cite{ahmed2024exeNet, ahmed2022less}. %, ensuring that raw samples and their augmented derivatives appeared only within the same split.

\begin{table}[t]
\centering
\caption{Properties of the BanglaSER dataset \cite{DAS2022108091}}
\label{table:banglaser_properties}
\begin{tabular}{lr}
\toprule
\textbf{Property} & \textbf{Value} \\
\midrule
No. of Speakers & 34 (17 male, 17 female) \\
Age Range & 19--47 years \\
No. of Utterances & 1,467 \\
No. of Classes & 5 (angry, happy, sad, surprise, neutral) \\
Samples per Class & 306 (neutral: 243) \\
Average Duration & 3--4 seconds \\
Audio Format & 16-bit PCM WAV \\
Validation Accuracy & 80.5\% (human raters) \\
\bottomrule
\end{tabular}
\end{table}

% To account for environmental interference, additive noise was injected into the raw waveform (\equationautorefname~\ref{eq:1}). Speaker variability was simulated through pitch perturbation by shifting the spectral distribution, where $\Delta$ corresponds to a semitone offset (\equationautorefname~\ref{eq:2}). Temporal variability was incorporated through rate modification (\equationautorefname~\ref{eq:3}) and through random temporal displacement (\equationautorefname~\ref{eq:4}), where $\tau$ is a random shift and $T$ is the signal duration. 

To account for environmental interference, additive Gaussian noise was injected into the raw waveform (\equationautorefname~\ref{eq:1}). Speaker variability was simulated through pitch perturbation, implemented by shifting the spectral distribution (\equationautorefname~\ref{eq:2}), where $\Delta$ denotes the semitone offset. Temporal variability was introduced using both rate modification (\equationautorefname~\ref{eq:3}) and random temporal displacement (\equationautorefname~\ref{eq:4}), with $\tau$ representing a random shift and $T$ the signal duration.

\begin{align}
    \tilde{x}(t) &= x(t) + \epsilon(t), \quad \epsilon(t) \sim \mathcal{N}(0, \sigma^2). \label{eq:1} \\
    \tilde{X}(f) &= X\!\left(f \cdot 2^{\Delta/12}\right) \label{eq:2} \\
    \tilde{x}(t) &= x(\alpha t), \quad \alpha \in [0.8,1.2] \label{eq:3} \\
    \tilde{x}(t) &= x\big((t+\tau) \bmod T\big) \label{eq:4}
\end{align}

To emphasize transient consonantal cues, percussive components were enhanced, while spectral equalization was applied using a low-pass filter (\equationautorefname~\ref{eq:5}), which attenuates higher-frequency energy to simulate channel-limited recordings.

\begin{equation}
\small
\tilde{X}(f) = H(f)X(f), \quad H(f) = \frac{1}{\sqrt{1+(f/f_c)^{2n}}}
\label{eq:5}
\end{equation}

Following augmentation, each waveform was transformed into both Chroma and Mel-frequency cepstral coefficients (MFCCs). The Chroma was computed from the short-time Fourier transform (STFT) magnitude $|S(t,f)|$ using a Mel filterbank $M$ (\equationautorefname~\ref{eq:6}). MFCCs were then obtained as the discrete cosine transform of the log-Mel spectrum (\equationautorefname~\ref{eq:7}). 

% \begin{align}
%     \text{MelSpec}(t,m) &= \sum_{f} M(m,f) \, |S(t,f)|^2 \label{eq:6}\\ 
%     \text{MFCC}(t,k) &= \sum_{m=1}^{M} \log(\text{MelSpec}(t,m)) \cos\!\left[\tfrac{\pi k}{M}(m-0.5)\right] \label{eq:7}
% \end{align}

\begin{center}
\small
\begin{align}
    \text{MelSpec}(t,m) &= \sum_{f} M(m,f) \, |S(t,f)|^2 \label{eq:6}\\
    \text{MFCC}(t,k) &= \sum_{m=1}^{M} \log(\text{MelSpec}(t,m)) \cos\!\left[\tfrac{\pi k}{M}(m-0.5)\right] \label{eq:7}
\end{align}
\end{center}

The final representation concatenates Chroma and MFCC features along the frequency axis, yielding a joint time--frequency descriptor that retains fine-grained spectral detail while compactly encoding the spectral envelope.

\subsection{Evaluated Architectures}

We evaluated a range of convolutional and transformer-based architectures using three spectrogram representations of speech signals: Mel-frequency cepstral coefficients (MFCC), Chroma features, and a fused representation combining both. Each feature type was preprocessed into fixed-size spectrogram images that served as inputs to the models. 
The selected architectures include DenseNet121 \cite{huang2017densely}, \backbone~\cite{cai2024efficientvitmultiscalelinearattention}, InceptionV3 \cite{szegedy2016rethinking}, MobileNetV2 \cite{sandler2019mobilenetv2invertedresidualslinear}, ResNet34, and ResNet50 \cite{he2016deepresiduallearning}. These models collectively span lightweight designs suitable for deployment in resource-constrained settings, such as MobileNetV2 and \backbone~\cite{ashmafee2023apple}, as well as deeper architectures capable of capturing complex hierarchical features, such as DenseNet121 and ResNet50 \cite{morshed2022Fruit}. These backbones were considered as they have consistently achieved state-of-the-art performance across a wide range of tasks and domains \cite{ahmed2025dexNet, raiyan2025hasper, fuad2025aqua20, herok2023cotton}.%, making them strong and reliable baselines for evaluating SER systems.
For each configuration, models were trained as end-to-end classifiers, receiving spectrogram images as input and producing emotion labels as output. 
% This setup enabled a systematic comparison across architectures and feature representations, allowing us to assess their relative effectiveness for Bangla speech emotion recognition.  

%%%%%%%%%%%%%%%%%%%%%%%%%%%%%%%%%%%%%%%%%%%%%%%%%%%%%%%%%%%%%%%%%%%%
\section{Result Analysis}
% This section presents a detailed evaluation of the proposed method, examining its overall performance, the impact of different feature representations, the behavior across emotion categories, its standing relative to existing approaches, and the structure of the learned feature space.

\subsection{Performance of different baseline architectures}  

We conducted extensive experiments on both the BanglaSER and SUBESCO datasets using three spectrogram representations: Chroma, MFCC, and a Combined variant integrating both. The evaluated architectures include several CNN-based and transformer-based backbones. \tableautorefname~\ref{table:performance_banglaser_subesco} summarizes the classification accuracy across all feature–model configurations.

Among the CNN architectures, DenseNet121 consistently produced the strongest results. Its best performance on SUBESCO reached 83.39\% with the Combined representation, reflecting its ability to model hierarchical spectral features effectively. ResNet34 also achieved competitive accuracy, obtaining 72.84\% on BanglaSER and 76.50\% on SUBESCO when using Combined features. In contrast, InceptionV3 yielded the lowest performance across both datasets, ranging from 53.24\% on BanglaSER (Chroma) to 65.17\% on SUBESCO (Combined), indicating limited suitability for Bangla spectrogram analysis. MobileNetV2, although efficient and lightweight, delivered only moderate results, with Combined accuracies of 68.06\% on BanglaSER and 67.29\% on SUBESCO.

In comparison, the transformer-based \backbone~model substantially outperformed all CNN baselines across datasets and feature types. It achieved state-of-the-art accuracies of 82.19\% on BanglaSER and 92.56\% on SUBESCO using the Combined representation, marking notable improvements over DenseNet121. These gains underscore the effectiveness of transformer architectures, particularly their ability to capture long-range temporal and spectral dependencies. The consistently higher performance across both low-resource (BanglaSER) and more controlled (SUBESCO) settings demonstrates the robustness and generalizability of our proposed approach.

\begin{table}[t]
\centering
% \caption{Performance comparison of different baseline architectures on BanglaSER and SUBESCO datasets}
\caption{Performance comparison of alternative backbone architectures evaluated within the proposed SpectroFusion-ViT framework on the BanglaSER and SUBESCO datasets.}
\label{table:performance_banglaser_subesco}

\resizebox{.98\columnwidth}{!}{%
\begin{tabular}{
    L{1.75cm}
    C{0.65cm} C{0.65cm} C{0.9cm} 
    C{0.65cm} C{0.65cm} C{0.9cm}
}
\toprule
& \multicolumn{3}{c}{\textbf{BanglaSER}} 
& \multicolumn{3}{c}{\textbf{SUBESCO}} \\
\cmidrule(lr){2-4} \cmidrule(lr){5-7}
\textbf{Architecture}
& \textbf{Chroma} & \textbf{MFCC} & \textbf{Combined}
& \textbf{Chroma} & \textbf{MFCC} & \textbf{Combined} \\
\midrule
Inception\_v3     & 53.24 & 55.91 & 58.39 & 59.42 & 62.07 & 65.17 \\
MobileNetv2      & 60.74 & 65.58 & 68.06 & 62.81 & 66.32 & 67.29 \\
ResNet50         & 64.38 & 67.12 & 69.30 & 63.55 & 65.41 & 67.58 \\
ResNet34         & 68.21 & 70.43 & 72.84 & 72.19 & 74.05 & 76.50 \\
DenseNet121      & 65.42 & 68.37 & 71.68 & \textbf{77.51} & 80.26 & 83.39 \\
\backbone & \textbf{70.55} & \textbf{79.65} & \textbf{82.19}
                 & 74.99 & \textbf{89.24} & \textbf{92.56} \\
\bottomrule
\end{tabular}
} % End of \resizebox
\end{table}

\subsection{Ablation Study}
We conducted an ablation study to evaluate the contribution of different spectrogram representations—Chroma, MFCC, and their combined variant across all backbone architectures. The corresponding results for both BanglaSER and SUBESCO are presented in \tableautorefname~\ref{table:performance_banglaser_subesco}. Across all models and datasets, the Combined representation consistently outperforms the individual feature types. This improvement reflects the complementary nature of Chroma and MFCC features. While Chroma captures harmonic and pitch-related structures, MFCC encodes broader spectral envelope information. Their integration therefore yields a more expressive time–frequency representation, enabling more reliable discrimination of emotional patterns in speech.

\begin{table}[b]
\centering
\caption{Class-wise precision, recall, and F1-score (\%) for BanglaSER and SUBESCO datasets}
\label{table:classwise_results}

% \resizebox{.98\columnwidth}{!}{%
\begin{tabular}{
    L{.9cm}
    C{0.8cm} C{0.65cm} C{1.1cm} 
    C{0.8cm} C{0.65cm} C{1.1cm}
}
\toprule
& \multicolumn{3}{c}{\textbf{BanglaSER}} 
& \multicolumn{3}{c}{\textbf{SUBESCO}} \\
\cmidrule(lr){2-4} \cmidrule(lr){5-7}
\textbf{Class}
& \textbf{Precision} & \textbf{Recall} & \textbf{F1-Score}
& \textbf{Precision} & \textbf{Recall} & \textbf{F1-Score} \\
\midrule
Angry   & 89 & 90 & 90 
        & 92 & 94 & 93\\
Happy   & 78 & 74 & 76 
        & 90 & 90 & 90\\
Neutral & 84 & 88 & 86 
        & 94 & 93 & 93\\
Sad     & 78 & 79 & 79 
        & 90 & 90 & 90\\
Surprise & 77 & 76 & 77 
        & 90 & 92 & 91\\
Disgust & - & - & - 
        & 88 & 87 & 88\\
Fear    & - & - & - 
        & 93 & 92 & 93\\

\bottomrule
\end{tabular}
% } % End of \resizebox
\end{table}

\subsection{Class-wise Analysis}

\tableautorefname~\ref{table:classwise_results} reports the per-class precision, recall, and F1-scores for both the BanglaSER and SUBESCO datasets. For BanglaSER, the model consistently achieved strong performance for \textit{Angry} (F1-score of 90\%) and \textit{Neutral} (86\%), indicating that these emotions are relatively well distinguished in Bangla speech. In contrast, \textit{Happy} and \textit{Surprise} were more challenging, with lower F1-scores of 76\% and 77\%, respectively; likely due to overlaps in pitch and prosodic patterns, which can reduce separability between positive and high-arousal emotions. The \textit{Sad} class showed moderate performance (79\%), suggesting some confusion with acoustically adjacent categories.

For SUBESCO, class-wise performance was generally higher and more balanced across emotions. The model achieved particularly strong results for \textit{Angry}, \textit{Fear}, and \textit{Neutral}, each with F1-scores above 93\%, reflecting clear emotional articulation and consistent recording quality. The \textit{Happy}, \textit{Sad}, and \textit{Surprise} categories also exhibited high accuracy ($approx$90\%), while \textit{Disgust} was slightly lower at 88\%, indicating that this emotion remains comparatively harder to distinguish.

These differences suggest that BanglaSER contains greater acoustic variability and emotional ambiguity, leading to lower class-wise separability. In contrast, the more controlled conditions and clearer emotional expressions in SUBESCO support higher per-class performance across all emotion categories.

\subsection{Comparison with state-of-the-art methods}

To further assess the effectiveness of the proposed approach, we compared our proposed pipeline \modelName~against existing state-of-the-art methods on the BanglaSER and SUBESCO datasets. The results are summarized in \tableautorefname~\ref{table:performance_analysis_with_sota_models}.

On BanglaSER, \modelName~achieved an accuracy of 82.19\%, exceeding the performance of Chakraborty \etal~\cite{chakraborty2022phase} (79.00\%) and Aziz \etal~\cite{aziz2023banglaSER} (78.00\%). Similarly, on SUBESCO, our method reached 92.56\%, outperforming Aziz \etal~\cite{aziz2023banglaSER} (90.00\%), the CNN-based system by Sultana \etal~\cite{9654185} (86.86\%), and Chakraborty \etal~\cite{chakraborty2022phase} (75.00\%).
These consistent improvements highlight the superior ability of \modelName~to model long-range temporal and spectral dependencies in emotional speech. 
% By leveraging transformer-based attention and a fused Chroma–MFCC representation, our approach establishes a new benchmark for Bangla speech emotion recognition on both BanglaSER and SUBESCO.

\begin{table}[h]
\centering
\caption{Comparison with state-of-the-art works on BanglaSER and SUBESCO datasets}
\label{table:performance_analysis_with_sota_models}
\begin{tabular}{l C{2.0cm} C{2.0cm}}
\toprule
& \multicolumn{2}{c}{\textbf{Accuracy (\%)}}\\
\cmidrule(lr){2-3}
\textbf{Architecture} & \textbf{BanglaSER} & \textbf{SUBESCO} \\ 
\midrule
Chakraborty \textit{et al.}\cite{chakraborty2022phase} & 79.00 & 75.00 \\ 
Sultana \textit{et al.}\cite{9654185} & -- & 86.86\\ 
Aziz \textit{et al.}\cite{aziz2023banglaSER} & 78.00 & 90.00 \\
Billah \textit{et al.}\cite{billah2024emotion} & -- & 90.14\\
Biswas \textit{et al.}\cite{Biswas2025} & -- & 90.50 \\
\modelName~(Ours) & \textbf{82.19} & \textbf{92.56}\\
\bottomrule
\end{tabular}
\end{table}

\subsection{Error Analysis}

We visualize the learned feature embeddings for the BanglaSER and SUBESCO datasets using t-SNE to analyze how emotions are organized in the latent space. This helps reveal patterns of inter-class overlap and intra-class variability that influence classification performance.

\begin{figure}[t]
\centering

\begin{subfigure}{0.49\linewidth}
    \centering
    \includegraphics[width=\linewidth]{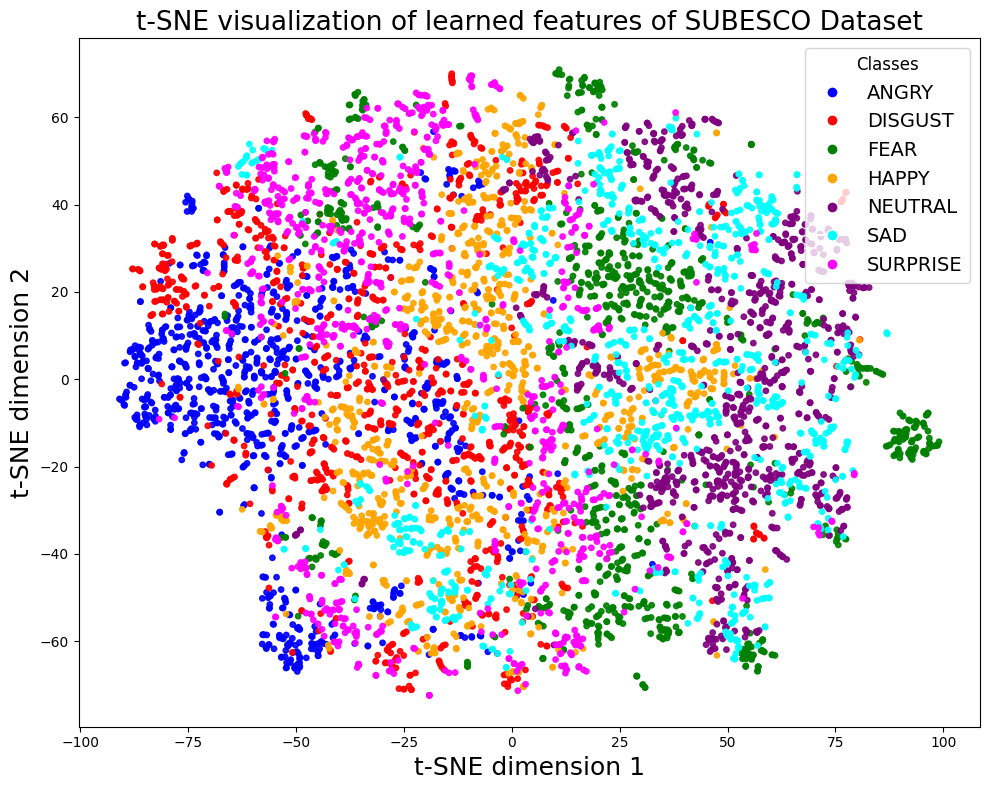}
    \caption{}
    \label{fig:subesco_tsne}
\end{subfigure}
\hfill
\begin{subfigure}{0.49\linewidth}
    \centering
    \includegraphics[width=\linewidth]{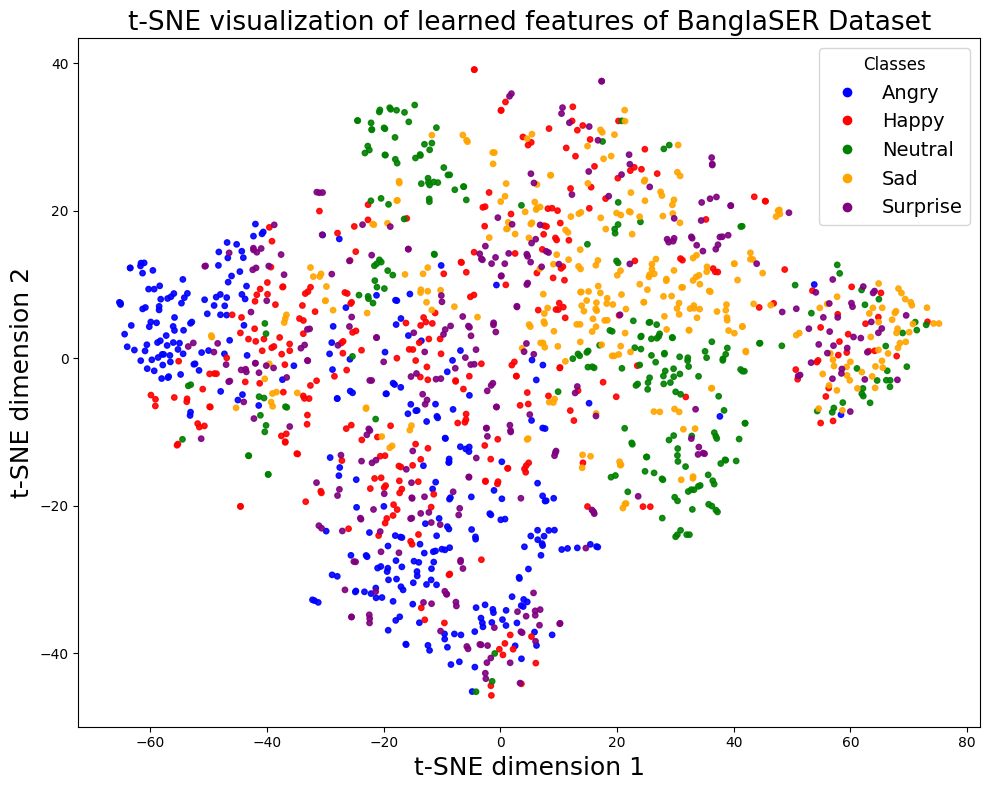}
    \caption{}
    \label{fig:banglaser_tsne}
\end{subfigure}

\caption{t-SNE visualizations of learned feature embeddings for (a) SUBESCO and (b) BanglaSER datasets.}
\label{fig:tsne_overall}
\end{figure}

For BanglaSER, which contains five emotion classes, the embeddings exhibit only partial clustering. \textit{Neutral} and \textit{Surprise} form relatively compact regions, whereas \textit{Happy}, \textit{Sad}, and \textit{Angry} show substantial overlap, as illustrated in \figureautorefname~\ref{fig:tsne_overall}\subref{fig:subesco_tsne}. This indicates that several emotions in Bangla speech share similar acoustic cues—particularly in prosody and energy contours—making them difficult for the model to distinguish. The broad spread of \textit{Angry} samples also reflects high intra-class variability, suggesting that speakers express this emotion with diverse acoustic patterns, which further reduces its separability.

In contrast, the SUBESCO dataset demonstrates much clearer clustering across the emotion classes. As shown in \figureautorefname~\ref{fig:tsne_overall}\subref{fig:banglaser_tsne}, \textit{Fear} and \textit{Surprise} form well-separated groups, and other classes such as \textit{Happy} and \textit{Sad} exhibit more distinct boundaries compared to BanglaSER. Although some minor overlap persists, the overall t-SNE distribution reveals tighter and more coherent clusters, indicating stronger emotional expressiveness and more consistent recording quality within SUBESCO.

These observations reinforce the performance trends reported earlier: BanglaSER contains greater acoustic ambiguity and higher inter-speaker variation, which increases classification difficulty. SUBESCO, on the other hand, provides richer emotional cues with clearer spectral distinctions, enabling the model to learn more discriminative embeddings and achieve higher inter-class separability.

% \begin{figure}[t]
% \centering

% \begin{subfigure}{0.9\columnwidth}
%     \centering
%     \includegraphics[width=\linewidth]{figures/tSNE/SUBESCO tSNE.png}
%     % \caption{t-SNE visualization of learned feature embeddings on the SUBESCO dataset.}
%     \label{fig:subesco_tsne}
% \end{subfigure}
% % \hfill

% \begin{subfigure}{0.9\columnwidth}
%     \centering
%     \includegraphics[width=\linewidth]{figures/tSNE/BanglaSER tSNE.png}
%     % \caption{t-SNE visualization of learned feature embeddings on the BanglaSER dataset.}
%     \label{fig:banglaser_tsne}
% \end{subfigure}

% \caption{t-SNE visualizations of learned feature embeddings for (a) SUBESCO and (b) BanglaSER datasets.}
% \label{fig:tsne_overall}
% \end{figure}

% \begin{figure}[t]
% \centering
% \includegraphics[width=0.45\textwidth]{figures/tSNE/SUBESCO tSNE.png}
% \caption{t-SNE visualization of learned feature embeddings on the SUBESCO dataset.}
% \label{fig:subesco_tsne}
% \end{figure}

% \begin{figure}[t]
% \centering
% \includegraphics[width=0.45\textwidth]{figures/tSNE/BanglaSER tSNE.png}
% \caption{t-SNE visualization of learned feature embeddings on the BanglaSER dataset.}
% \label{fig:banglaser_tsne}
% \end{figure}

\section{Conclusion}\label{conclusion}

% For a better understanding of human speech, accurate identification of emotion is important by minimizing misclassifications and maximizing performance. In this work, we introduced a lightweight vision transformer architecture with a self-attention mechanism to effectively learn discriminative features and capture patterns from spectrogram representations of audio. We also applied several data preprocessing strategies, including data augmentation, to enhance robustness and generalization. Our experimental results demonstrate strong performance, achieving state-of-the-art accuracy on benchmark datasets. For future work, the system can be further optimized for improved accuracy and extended to real-time applications on resource-constrained devices.  
% Accurately identifying emotions from speech is essential for improving human–computer interaction systems and enabling more adaptive and context-aware applications. 
This work presents \modelName, a lightweight vision transformer–based framework that leverages self-attention to learn discriminative temporal–spectral patterns from speech. The model integrates a fused Chroma–MFCC representation and a comprehensive augmentation pipeline to enhance robustness and generalization under diverse acoustic conditions.
Our experimental analysis demonstrates that \modelName~achieves state-of-the-art performance on both BanglaSER and SUBESCO, outperforming several strong CNN and transformer baselines while maintaining low computational complexity. These results highlight the effectiveness of combining harmonic–spectral fusion with efficient transformer architectures for speech emotion recognition.
In the future, the framework can be further optimized for improved accuracy, extended to multilingual or cross-corpus scenarios, and deployed in real-time systems on resource-constrained devices. Incorporating additional modalities such as text or facial cues may also further enhance emotion recognition in practical environments.

\bibliographystyle{ieeetr}
\bibliography{citations}

\end{document}